\documentclass[11pt]{article}

\def\be{\begin{equation}}
\def\ee{\end{equation}}
\def\bea{\begin{eqnarray}}
\def\eea{\end{eqnarray}}

\usepackage{a4wide}
\usepackage{bm}
\usepackage{cite}

\begin{document}

\author{Bruno Boisseau \thanks{E-mail: Bruno.Boisseau@lmpt.univ-tours.fr}
\quad  Bernard Linet  \thanks{E-mail: Bernard.Linet@lmpt.univ-tours.fr} \\
\small Laboratoire de Math\'ematiques et Physique Th\'eorique \\
\small CNRS/UMR 7350, F\'ed\'eration Denis Poisson (FR2964) \\
\small Universit\'e Fran\c{c}ois Rabelais, 37200 TOURS, France}

\title{\bf Electrostatics in a simple wormhole revisited}

\date{}

\maketitle

\thispagestyle{empty}

\begin{abstract}
The electrostatic potential generated by a point charge at rest in a simple static, spherically symmetric wormhole is given
in the form of series of multipoles and in closed form. The general potential which is  physically acceptable depends on a parameter
due to the fact that the monopole solution is arbitrary. When the wormhole has
$Z_2$-symmetry, the potential is completly determined.
The calculation of the electrostatic self-energy and of the self-force is performed in all cases considered.
\end{abstract}

\section{Introduction}

The determination in the form of series of multipoles of the electrostatic potential generated by a point charge at rest in 
some static, spherically symmetric wormholes has been a subject of investigation recently.
The main purpose of all these works \cite{khus1, khus2, krasnikov, khus3, beskin} is to determine the electrostatic self-force acting
on the point charge within this multipole formalism. We note that the self-force for the scalar field has been also considered 
\cite{bezerra, popov}.

The determination in closed form of the electrostatic potential is interesting on its own. In principle, it is easier to calculate 
the electrostatic self-energy of a point charge from the electrostatic potential in closed form rather than given in the form 
of series of multipoles. It may be practical to have an expression easy to handle for some calculations.

This paper is devoted in the resolution of the electrostatic potential in a particularly simple wormhole.
The spacetime representing the simplest kind of static, spherically symmetric wormhole has been popularized by Morris and Thorne 
in a pedagogical paper \cite{morris} but it had been already proposed by Ellis \cite{ellis} and Bronikov \cite{bronikov}. 
The metric describing this spacetime can be written
\be \label{m}
ds^2=-dt^2+dl^2+(l^2+w^2)(d\vartheta^2+\sin^2\vartheta d\varphi^2)
\ee
in the coordinate system $(t,l,\vartheta ,\varphi )$, with $-\infty < l < \infty$, where $w$ is a constant. The space-time of 
this wormhole is regular everywhere. The throat is located at $l=0$. It admits two distinct, asymptotical spacetimes which are 
obtained in the limit where $l\rightarrow \pm \infty$ in metric (\ref{m}); these are two Minkowskian spacetimes. One can find 
an overview of wormholes in the book of Visser \cite{visser}.

We consider a charged particle at rest in this spacetime, assuming that the electromagnetic field $F_{\mu \nu}$ 
is sufficiently weak so that its gravitational effect is negligible. In this case, the electromagnetic field satisfies the Maxwell 
equations in the background metric (\ref{m}). The charge density $\rho$ of a point charge $e$ located at $l=l_0$, $\vartheta = \vartheta_0$ 
and $\varphi =\varphi_0$ is given by
\be \label{dc}
\rho (l,\vartheta ,\varphi )=\frac{e}{\sqrt{-g}}\delta (l-l_0)\delta (\vartheta -\vartheta_0)\delta (\varphi -\varphi_0)
\ee
where $g$ is the determinant of metric (\ref{m}).

The electrostatic potential $V$, defined from the electric field $F_{0i}$ by $F_{0i}=-\partial_i V$, obeys the electrostatic equation 
\be \label{eV}
\frac{\partial^2}{\partial l^2}V+\frac{2l}{l^2+w^2}
\frac{\partial}{\partial l}V+\frac{1}{(l^2+w^2)} \left[ \frac{1}{\sin \vartheta}\frac{\partial}{\partial \vartheta}
\left( \sin \vartheta \frac{\partial}{\partial \vartheta}V\right) +\frac{1}{\sin^2 \vartheta}\frac{\partial^2 }{\partial \varphi^2}V \right] =-4\pi \rho .
\ee
To obtain a physical solution to equation (\ref{eV}), we must add the asymptotic condition that the associated electric field
$F_{0i}$ vanishes at the two spatial infinities, i.e. as $l\rightarrow \pm \infty$, because only the electric field has a direct
physical meaning.

A solution to equation (\ref{eV}) was given for the first time in the form of series of multipoles by Khusnutdinov and 
Bakhmatov \cite{khus1}. Recently, Beskin {\em et al} \cite{beskin} took up the issue of the multipole expansion. 

However, one of us in a previous preprint \cite{linet1}  has solved the electrostatic equation (\ref{eV}) in isotropic
coordinates giving an explicit expression of the potential $V$ in closed form. A part of this paper is devoted to clarify the relation
between these two forms of the solution. Firstly, we find again the multipole expansion of $V$ in a simple and synthetic manner. 
Secondly, we enlarge the result of the preprint \cite{linet1} to a wormhole in $n+1$ dimensions, $n\geq 3$, described by a direct 
generalisation of metric (\ref{m}). To facilitate the reading of the paper, we give the complete proof in an appendix.
In the case $n=3$, we express the explicit expression of $V$ in terms of the radial coordinate $l$. This allows us 
to compare easily with the multipole expansion.

We then prove that the two determinations of $V$ coincide. However, this above mentioned potential $V$ is a particular solution 
to equation (\ref{eV}). In fact, we can add the monopole solution which satisfies the boundary condition about 
the electric field at $l\rightarrow \pm \infty$.
Assuming that the electrostatic potential must be symmetric in $l$ and $l_0$, the general solution $V_{\lambda}$
to equation (\ref{eV}), physically acceptable, depends on an arbitrary parameter $\lambda$.

We also consider the spacetime resulting from an identification according to the $Z_2$ symmetry, described by metric (\ref{m}) with $l\geq 0$.
We determine the electrostatic potential $V_Z$ generated by a point charge in this case. In this case it is unique.

We find the explicit expression $W$ of the electrostatic self-energy for the cases $V$, $V_{\lambda}$ and $V_Z$
by using the usual procedure. By deriving $W$, we then deduce the electrostatic self-force which has only a radial component.

The paper is organized as follows. In section 2, we give the expressions of the radial functions of the multipole solutions.
We express the electrostatic potential in the form of series of multipoles in section 3. We discuss in section 4 the general potential
 which is physically acceptable. We determine in section 5
the electrostatic potential in closed form and we prove that it coincides with the above multipole expansion.  We obtain the electrostatic potential
in the wormhole with the $Z_2$ symmetry in section 6. We perform the calculation of the electrostatic self-energy and of the self-force
in section 7 in all cases considered. We add some concluding remarks in section 8.
 
\section{Radial functions of the multipole solutions}

To simplify the writing of the formulas, we introduce the dimensionless variables $x$ and $a$ 
\be \label{xl}
x=\frac{l}{w} , \quad a=\frac{l_0}{w} ,
\ee
and the potential without dimension $A$ by putting
\be \label{VA}
A(x,\vartheta ,\varphi )=\frac{w}{e}V(x,\vartheta ,\varphi ) .
\ee
The electrostatic equation (\ref{eV}) becomes
\bea \label{eA}
& & \frac{\partial^2}{\partial x^2}A+\frac{2x}{x^2+1}\frac{\partial}{\partial x}A+\frac{1}{x^2+1}\left[ \frac{1}{\sin \vartheta}
\frac{\partial}{\partial \vartheta}\left( \sin \vartheta \frac{\partial}{\partial \vartheta}A \right) +
\frac{1}{\sin^2\vartheta}\frac{\partial^2}{\partial \varphi^2}A \right] = \nonumber \\
& & \qquad -\frac{4\pi}{x^2+1}\delta (x-a)\delta (\cos \vartheta -\cos \vartheta_0)\delta (\varphi -\varphi_0) .
\eea

We will seek the multipole solutions to the corresponding homogeneous equation 
\be \label{eAh}
 \frac{\partial^2}{\partial x^2}A+\frac{2x}{x^2+1}\frac{\partial}{\partial x}A+\frac{1}{x^2+1}\left[ \frac{1}{\sin \vartheta}
\frac{\partial}{\partial \vartheta}\left( \sin \vartheta \frac{\partial}{\partial \vartheta}A \right) +
\frac{1}{\sin^2\vartheta}\frac{\partial^2}{\partial \varphi^2}A \right] = 0 .
\ee
 They are obtained by separation of variables
$x$, $\vartheta$ and $\varphi$. In spherical symmetry, it is well known that this is performed by using the spherical harmonics
$Y_{n}^{m}(\vartheta ,\varphi )$, $n=0,1,\dots $ with $| m | \leq n$. The multipole solutions have thus the form
\be \label{sm}
A_{n}^{m}(x,\vartheta ,\varphi )= A_n(x) Y_{n}^{m}(\vartheta ,\varphi ) .
\ee
By substituting form (\ref{sm}) into equation (\ref{eAh}), we find that each radial function
$A_n$ satisfies the differential equation
\be \label{smh}
\frac{d^2A_n}{dx^2}+\frac{2x}{x^2+1}\frac{dA_n}{dx}-\frac{n(n+1)A_n}{x^2+1} =0 .
\ee

The coefficients of the linear differential equation (\ref{smh}) are regular everywhere and therefore the vector space of the solutions
consists of smooth functions. The solutions have two types of asymptotic behavior: $x^n$ and $1/x^{n+1}$ as $x\rightarrow \infty$.
We have to define two linearly independent solutions corresponding to the two asymptotic behaviors that we denote respectively $f_n$
and $g_n$, the latter been fixed up to a multiplicative constant.

We introduce the variable $z$ by setting
$$
z=\imath x ,\quad {\rm or}\quad x=-\imath z ,
$$
where $\imath$ is the unit imaginary number.
The differential equation (\ref{smh}) expressed in $z$ becomes the Legendre differential equation whose the theory 
in the complex plane is well known \cite{olver}. The two fundamental solutions in the complex plane are denoted 
$P_n(z)$ and $Q_n(z)$. The $P_n(z)$ are polynomials in $z$ of degree $n$ whose powers of $z$ are either odd or even.
The functions $Q_n(z)$ can be written
$$
Q_n(z)=\frac{1}{2}P_n(z)\ln \left( \frac{z+1}{z-1}\right) +w_{n-1}(z) 
$$
where the $w_{n-1}(z)$ are polynomials of degree $n-1$, with $w_{-1}=0$, whose powers of $z$ are either odd or even, 
given for instance by
$$
w_{n-1}(z)=-\sum_{q=1}^{n} \frac{1}{q} P_{q-1}(z)P_{n-q}(z) .
$$

The Wronskian of these solutions is
$$
{\cal W}(P_n,Q_n)=-\frac{1}{z^2-1} .
$$
They admit respectively the asymptotic behaviors
$$
P_n(z)\sim \frac{2^n\Gamma (n+\frac{1}{2}) z^n}{\sqrt{\pi}\Gamma (n+1)} \quad {\rm as} \quad z\rightarrow \infty ,
$$
$$
Q_n(z)\sim \frac{\sqrt{\pi}\Gamma (n+1)}{2^{n+1}\Gamma (n+\frac{3}{2}) z^{n+1}} \quad {\rm as} \quad z\rightarrow \infty .
$$

The above results suggest immediately that one can take
\be \label{df}
f_n(x)=\frac{1}{\imath^n}P_n(\imath x) 
\ee
as a real solution to the differential equation (\ref{smh}) with the asymptotic behavior
\be \label{af}
f_n(x)\sim \frac{2^n\Gamma (n+\frac{1}{2}) x^n}{\sqrt{\pi}\Gamma (n+1)} \quad {\rm as} \quad x\rightarrow \infty .
\ee

On the other hand, we cannot take $Q_n(\imath x)/i^{n-1}$ as the other solution $g_n$ because the logarithmic term
in $Q_n(\imath x)$ yields a problem at $x=0$. Indeed, the expression of $Q_n(z)$ for $z=\imath x$ is
$$
Q_n(\imath x)=\frac{1}{2}P_n(\imath x)\ln \left( \frac{\imath x+1}{\imath x-1}\right) +w_{n-1}(\imath x) .
$$
But we have the identity
$$
\arctan \frac{1}{x} =\frac{i}{2}\ln \left( \frac{\imath x+1}{\imath x-1}\right) ,\quad x\neq 0 ,
$$
and therefore
$$
Q_n(\imath x)=\frac{P_n(\imath x)}{\imath}\arctan \frac{1}{x}+w_{n-1}(\imath x) ,\quad x\neq 0 .
$$
There exists a discontinuity of the function $Q_n(\imath x)$ at $x=0$.

Considering the above expression, we define $g_n(x)$ for positive $x$ by putting
\be \label{dgp}
g_n(x)=(-1)^n\frac{Q_n(\imath x)}{\imath^{n-1}} , \quad x>0 .
\ee
The factor $(-1)^n$ is introduced for later so that the Wronskian of $f_n$ and $g_n$ does not depend on $n$.
Taking into account the identity
$$
\arctan \frac{1}{x}=\frac{\pi}{2}-\arctan x, \quad x>0,
$$
we write $g_n(x)$ for $x>0$ under the form

\be \label{dg}
g_n(x)=(-1)^n\left[ f_n(x)\left( \frac{\pi}{2}-\arctan x\right)+v_{n-1}(x)\right] 
\ee
where the $v_{n-1}(x)=w_{n-1}(\imath x)/\imath^{n-1}$ are real polynomials in $x$ of degree $n-1$.

But of course, we can now extend expression (\ref{dg}) for $x\leq 0$. Thus, 
expression (\ref{dg}) is a real solution to the differential equation (\ref{smh}) 
whose asymptotic behavior is
\be \label{ag}
g_n(x)\sim \frac{\sqrt{\pi}\Gamma (n+1)}{2^{n+1}\Gamma (n+\frac{3}{2})x^{n+1}} \quad {\rm as} \quad x\rightarrow \infty .
\ee

Taking into account the identity
$$
\arctan \frac{1}{x}=-\frac{\pi}{2}-\arctan x , \quad x<0 ,
$$
we note that an equivalent way to express (\ref{dg}) for negative  $x$ is the formula
\be \label{dgn}
g_n(x)=(-1)^n\left( \frac{Q_n(\imath x)}{\imath^{n-1}} +\pi \frac{P_n(\imath x)}{\imath^n}\right) , \quad x<0 .
\ee

We have therefore found two regular solutions $f_n$ and $g_n$ to the differential equation (\ref{smh}) which are linearly independent, with the Wronskian
\be \label{wfg}
{\cal W}(f_n,g_n)=-\frac{1}{1+x^2} ,
\ee
and which have respectively the asymptotic behaviors (\ref{af}) and (\ref{ag}) as $x\rightarrow \infty$.

We may also consider the asymptotic behavior as $x\rightarrow -\infty$ of the solutions to the differential equation
(\ref{smh}). The form of this latter shows that starting from a solution we obtain a new solution by remplacing $x$ by $-x$.
So we can define the solution $g_n(-x)$. It has the asymptotic behavior $1/(-x)^{n+1}$ as $x\rightarrow -\infty$.

By using the parity properties of polynomials $f_n$ and $v_{n-1}$, it can be easily shown that
\be \label{egn}
g_n(-x)=\pi f_n(x)-(-1)^ng_n(x) ,
\ee
and according to (\ref{wfg}) we have
\be \label{45}
{\cal W}(g_n(-x),g_n(x))=-\frac{\pi}{1+x^2} .
\ee

\section{Electrostatic potential in series of multipoles}

The form of the multipole solutions (\ref{sm}) suggests to seek the solution $A$ to the electrostatic equation (\ref{eA})
in the form of series of multipoles.
By expressing the second member of equation (\ref{eA}) in terms of spherical harmonics by the well known formula
\be \label{42} 
\delta (\cos \vartheta -\cos \vartheta_0)\delta \varphi -\varphi_0) =
\sum_{n=0}^{\infty}\sum_{m=-n}^{n}Y_{n}^{m*}(\vartheta_0,\varphi_0)Y_{n}^{m}(\vartheta ,\varphi ) ,
\ee
we see easily that the potential $A$ has the form
\be \label{41}
A(x,\vartheta ,\varphi )=\sum_{n=0}^{\infty}\sum_{m=-n}^{n}A_n(x,a)Y_{n}^{m*}(\vartheta_0,\varphi_0)Y_{n}^{m}(\vartheta ,\varphi ) ,
\ee
where $*$ denotes the complex conjugate.
By reporting (\ref{41}) into (\ref{eA}), we find that each component $A_n$ has to satisfy the differential equation 
\be \label{43}
\frac{d^2A_n}{dx^2}+\frac{2x}{x^2+1}\frac{dA_n}{dx}-\frac{n(n+1)A_n}{x^2+1} =-\frac{4\pi}{1+x^2}\delta (x-a) .
\ee

To solve equation (\ref{43}) we must add the asymptotic behaviors of $A_n$ as $x\rightarrow \pm \infty$.
As stated in the introduction, we require that the asymptotic electric field vanishes. Taking into account the asymptotic behaviors of the radial functions, 
it is sufficient to require
\be \label{4a}
\lim_{x\rightarrow -\infty}A_n(x,a)=0 ,\quad \lim_{x\rightarrow \infty}A_n(x,a)=0 ,
\ee
defining thus a solution $A$ to the electrostatic equation (\ref{eA}).

We now use the results about the Green function of a self-adjoint differential equation, subject to asymptotic behaviors (\ref{4a}), given
in the appendix A. We apply for the case
$$
p(x)=1+x^2 ,\quad q(x)=-n(n+1) , \quad c=-4\pi .
$$
The homogeneous equation associated to equation (\ref{43}) is identical to equation (\ref{smh}) which admits the two
regular, linearly independent solutions $g_n(-x)$ and $g_n(x)$, verifying the asymptotic behaviors
\be \label{4l}
\lim_{x\rightarrow -\infty} g_n(-x)=0 , \quad \lim_{x\rightarrow \infty} g_n(x)=0 
\ee
as shown in the previous section. Taking into account (\ref{45}), we put
\be \label{46}
A_{n}^{(i)}(x)=2g_n(-x) , \quad A_{n}^{(e)}(x)=2g_n(x) .
\ee
We thus rewrite the Green function (\ref{B9}) in terms of the functions $g_n$ in the form
\be \label{4b}
A_n(x,a)=4\theta (a-x)g_n(-x)g_n(a)+4\theta (x-a) g_n(-a)g_n(x) ,
\ee
where $\theta$ is the Heaviside step function.

To see that our result coincides with the one of Khusnutdinov and Bakhmatov \cite{khus1}, we must replace
our radial functions $g_n(x)$ and $g_n(-x)$ by their expressions in terms of the Legendre functions 
according to the sign of $x$. For $g_n(x)$, we have (\ref{dgp}) and (\ref{dgn}) and from (\ref{egn}) we get
\be \label{dgpn}
g_n(-x)=\pi \frac{P_n(\imath x)}{\imath^n}-\frac{Q_n(\imath x)}{\imath^{n-1}} , \quad x>0 ,
\ee
\be \label{dgnn}
g_n(-x)=-\frac{Q_n(\imath x)}{\imath^{n-1}} , \quad x<0 .
\ee

It will be useful later to know the asymtotic behavior of the potential $A$ as $l\rightarrow \infty$, or equivalently the electric flux $\Phi_{(\infty )}$
at the infinity $l\rightarrow \infty$. According to (\ref{ag}) and (\ref{4b}), the leading term in $1/x$ in expression (\ref{41}) of 
the potential $A$ can only come
from the term $4g_0(-a)g_0(x)$. Let us note that
\be \label{fg0}
f_0(x)=1 , \quad g_0(x)=\frac{\pi}{2}-\arctan x ,
\ee
and that $Y_{0}^{0}=1/\sqrt{4\pi}$. So we have
\be \label{af0}
A(x,\vartheta ,\varphi ) \sim \frac{4}{x}\frac{1}{4\pi}\left( \frac{\pi}{2}+\arctan a\right) \quad {\rm as}\quad x\rightarrow \infty .
\ee
Expression (\ref{af0}) yields the electric flux $\Phi_{(\infty)}$ at the infinity $l\rightarrow \infty$ given by
\be \label{fA}
\Phi_{(\infty)}=4\pi e\left( \frac{1}{2}+\frac{1}{\pi}\arctan \frac{l_0}{w}\right) .
\ee

\section{ Physically acceptable general solution}

Solution (\ref{4b}) allows us to write an electrostatic potential (\ref{41}) satisfying the boundary condition about the associated electric field. 
This is not the general solution to the electrostatic equation (\ref{eA}) because we can add some homogeneous solutions $A_{n}^{m}$ given by (\ref{sm}).
However, only the electric field associated to the monopole solution $A_{0}^{0}$, given by
$$
A_{0}^{0}(x)=\frac{1}{\sqrt{4\pi}} f_0(x) , \quad A_{0}^{0}(x)=\frac{1}{\sqrt{4\pi}}g_0(x) ,
$$
vanishes both at the two spatial infinities, i.e. $x\rightarrow \pm \infty$. 

In consequence, the  physically acceptable general solution to the electrostatic equation (\ref{eA}) has the form 
\be \label{Agg}
A_{gen}(x,\vartheta ,\varphi )=  A(x,\vartheta ,\varphi )+\lambda (a)g_0(x) ,
\ee
where $\lambda$ is an arbitrary function of $a$. It cannot be determined by the study of the electrostatic equation (\ref{eA})
without another condition. Since the source term is symmetric in $x$ and $a$, we may take the assumption 
that the general solution must be symmetric in $x$ and $a$. This determines the function $\lambda$ in
the form $\lambda (a)=\lambda g_0(a)$ where $\lambda$ is now an arbitrary constant. So we put
\be \label{Ag0}
A_{\lambda}(x,\vartheta ,\varphi )=  A(x,\vartheta ,\varphi )+\lambda g_0(a)g_0(x) .
\ee

For the general solution $A_{\lambda}$ given by (\ref{Ag0}), we obtain the electric flux $\Phi_{\lambda (\infty )}$ 
by adding the monopole contribution to the electric flux (\ref{fA}). We find 
\be \label{fAg}
\Phi_{\lambda (\infty)}=4\pi e\left( \frac{1}{2}+\frac{1}{\pi}\arctan \frac{l_0}{w}\right) +4\pi e \lambda 
\left( \frac{\pi}{2}-\arctan \frac{l_0}{w}\right).
\ee

Beskin {\em et al} \cite{beskin} suggested the need to impose the additional condition that the electric flux
at the infinity $l\rightarrow \infty$ is exactly $4\pi e$
as expected in the usual cases at the spatial infinity. From expression (\ref{fAg}), we obtain this condition for $\lambda =1/\pi$. 
However, this choice favors one of the two infinities.

\section{Electrostatic potential in closed form}

We give in appendix B the electrostatic potential in a wormhole described by  metric (\ref{m}) written in 
$n+1$ dimensions, $n\geq 3$. Its explicit expression is found in isotropic coordinates of this metric. In this section, 
we recall only the basic formulas of appendix B useful to understand the expression of the electrostatic potential $V$.

We firstly write metric (\ref{m}) in isotropic form by introducing the radial coordinate $r$ by
\be \label{ir}
r=\frac{1}{2}\left( l+\sqrt{l^2+w^2}\right) , \quad {\rm or}\quad l=r-\frac{w^2}{4r}  ,
\ee
and then metric (\ref{m}) takes the form
\be \label{im}
ds^2=-dt^2+\left( 1+\frac{w^2}{4r^2}\right)^2 \left( dr^2+r^2(d\vartheta^2+\sin^2\vartheta d\varphi^2)\right) 
\ee
with $r>0$. The position $r_0$ of the point charge is given by
\be \label{ir0}
r_0=\frac{1}{2}\left( l_0+\sqrt{l_{0}^{2}+w^2}\right) , \quad {\rm or} \quad l_0=r_0-\frac{w^2}{4r_0}  .
\ee

Due to the spherical symmetry, without lost of generality,
we write down the expression of the electrostatic potential in metric (\ref{im}) for a point charge $e$ located at $r=r_0$ 
and $\vartheta_0=0$ by using (\ref{AVex}) with (\ref{Aekg}) for $n=3$. We have the form
\be \label{iVex} 
V(r,\vartheta )=\frac{F(s)}{\displaystyle \left( r+\frac{w^2}{4r}\right) \left( r_0+\frac{w^2}{4r_0}\right)} ,
\ee
where $s$ defined by (\ref{Ais}) reduces to
\be \label{is}
s=\frac{w^2(r^2-2rr_0\cos \vartheta +r_{0}^{2})}{4(r^2+w^2/4)(r_{0}^{2}+w^2/4)} , \quad 0\leq s\leq 1 ,
\ee
and where $F(s)$ is given by (\ref{AVf}) for $n=3$ in the form

\be \label{F}
F(s)=\frac{ew}{2\sqrt{s(1-s)}}\left[ 1+\frac{1}{\pi}\left( \arcsin (1-2s)-\frac{\pi}{2}\right) \right] .
\ee
Expression (\ref{iVex}) with (\ref{F}) yields an electrostatic potential $V$ which is regular everywhere except at the position of the point charge, 
characterized by $s=0$.

In this paper, we add the determination of the electrostatic potential (\ref{iVex}) in terms of $l$ and $l_0$. 
By substituting (\ref{ir}) and (\ref{ir0}) into (\ref{iVex}), we have firstly
\be \label{s}
1-2s=\frac{ll_0+w^2\cos \vartheta}{\sqrt{l^2+w^2}\sqrt{l_{0}^{2}+w^2}} ,
\ee
and finally we obtain 
\be \label{Vex} 
V(l,\vartheta )=\frac{e}{\sqrt{l^2-2ll_0\cos \vartheta +l_{0}^{2}+w^2\sin^2\vartheta}} \left[ \frac{1}{2}+
\frac{1}{\pi}\arcsin \left( \frac{ll_0+w^2\cos \vartheta}{\sqrt{l^2+w^2}\sqrt{l_{0}^{2}+w^2}}\right) \right] .
\ee
Of course, potential (\ref{Vex}) is a solution to electrostatic equation (\ref{eV}). The associated electric field
vanishes as $l\rightarrow \pm \infty$. We point out that expression (\ref{Vex}) of 
the electrostatic potential $V$ is much simpler in these coordinates.

We now determine the electric flux $\Phi_{(\infty)}$ as $l\rightarrow \infty$ corresponding to solution (\ref{Vex}). 
We find immediately 
\be \label{fV}
\Phi_{(\infty)} = 4\pi e \left[ \frac{1}{2}+\frac{1}{\pi}\arcsin \left( \frac{l_0}{\sqrt{l_{0}^{2}+w^2}} \right) \right] .
\ee
With the help of the identity
$$
\arctan a=\arcsin \left( \frac{a}{\sqrt{a^2+1}}\right) ,
$$
we see that the electric flux (\ref{fV}) coincides with the one of the solution in series of multipoles given by (\ref{fA}).

Since the only possible difference between two physical solutions is proportional to the monopole solution $g_0(l/w)$,
we therefore conclude that the potential in closed form and the previous potential in series of multipoles can be identified.
Of course, we may add to $V$ the monopole solution for obtaining the general solution $V_{\lambda}$ which is physically acceptable
\be \label{Vg0}
V_{\lambda}(l,\vartheta ,\varphi )=V(l,\vartheta ,\varphi )+\lambda \frac{e}{w}g_0(l_0/w)g_0(l/w) .
\ee

\section{Wormhole metric with the $Z_2$ symmetry}

We have considered until now that the spacetime is described by metric (\ref{m}) with $-\infty <l<\infty$. However,
we can impose the $Z_2$ symmetry by the topological identification of the points $(l,\vartheta ,\varphi )$ and
$(-l,\vartheta ,\varphi )$ of this spacetime. Such a $Z_2$ symmetry for massive wormhole models have been
proposed as black hole foils \cite{damour}.

The metric describing this new spacetime is again metric (\ref{m}) but with $l\geq 0$. However, the
physical laws in the background metric (\ref{m}) must satisfy some boundary conditions at $l=0$ compatible
with the $Z_2$ symmetry.
In electrostatics, as already noticed in \cite{damour}, the radial component of the electric field must vanish at $l=0$.

We take a point charge $e$ located at $l=l_0$ and $\vartheta =0$ with $l_0>0$. The electrostatic potential $V_Z$ 
satisfies the electrostatic equation (\ref{eV}) for $l\geq 0$ with the boundary condition
\be \label{Z1}
\frac{\partial V_Z}{\partial l}\bigg|_{l=0} = 0.
\ee

It is easy to determine $V_Z$ by introducing, before the
topological identification, a point charge $e$ located at $l=-l_0$ and $\vartheta =0$. We get from (\ref{Vex}) the expression
\bea \label{Z2}
& & V_Z(l,\vartheta )=\frac{e}{\sqrt{l^2-2ll_0\cos \vartheta +l_{0}^{2}+w^2\sin^2\vartheta}} \left[ \frac{1}{2}+
\frac{1}{\pi}\arcsin \left( \frac{ll_0+w^2\cos \vartheta}{\sqrt{l^2+w^2}\sqrt{l_{0}^{2}+w^2}}\right) \right] +  \nonumber \\
& & \quad \frac{e}{\sqrt{l^2+2ll_0\cos \vartheta +l_{0}^{2}+w^2\sin^2\vartheta}} \left[ \frac{1}{2}+
\frac{1}{\pi}\arcsin \left( \frac{-ll_0+w^2\cos \vartheta}{\sqrt{l^2+w^2}\sqrt{l_{0}^{2}+w^2}}\right) \right] ,
\eea
which does satisfy condition (\ref{Z1}).
We emphasize that solution (\ref{Z2}) is unique because we cannot add the monopole solution $g_0(l/w)$ which does not verify the 
boundary condition (\ref{Z1}).

By taking the limit of expression (\ref{Z2}) as $l\rightarrow \infty$, we can calculate the electric flux $\Phi_{Z(\infty )}$
at infinity. We find thereby
\be \label{Z3}
\Phi_{Z(\infty )}=4\pi e ,
\ee
as expected from the Gauss theorem since there is no electric flux through the sphere $l=0$ by virtue of (\ref{Z1}).

\section{Electrostatic self-force}

Besides a divergent electric field at the position of the point charge, there is a finite electric field which may be considered
as an external field. It exerts on the point charge an electrostatic force. So,
we have firstly to exhibit the singular part of the potential which leads to the divergent electric field.

We start with the electrostatic potential (\ref{iVex})  written in isotropic coordinates. According to (\ref{AapV}) for $n=3$, 
the singular part $V_{sing}$ of this potential is simply
\be \label{61}
V_{sing}(x)=\frac{e}{\sqrt{1+w^2/4r^{2}}\sqrt{1+w^2/4r_{0}^{2}}}\frac{1}{[\Gamma (x,x_0)]^{1/2}} .
\ee
Since the function $F$ given by (\ref{F}) admits the expansion
\be \label{62}
F(s) \sim \frac{ew}{2} \left( \frac{1}{\sqrt{s}}-\frac{2}{\pi}+\frac{1}{2}\sqrt{s}\right)  \quad {\rm as}\quad s\rightarrow 0 ,
\ee
and taking into account (\ref{iVex}), we obtain 
\be \label{63}
V(x)\sim V_{sing}(x)-\frac{ew}{\pi \displaystyle \left( r+\frac{w^2}{4r}\right) \left( r_0+\frac{w^2}{4r_0}\right)}
\quad {\rm as} \quad x\rightarrow x_0 .
\ee

The regular term in the neighbourhood of $x_0$ can be seen as resulting from the difference between $V$ and the
elementary solution in the Hadamard sense to the electrostatic equation. The method of regularization from the
electromagnetic Green function in an ultrastatic spacetime leads to simply subtract $V_{sing}$ to $V$ \cite{casals}.

So the electrostatic self-force derives from the self-energy $W$ as follows
$$
f^i(x_0)=-\frac{\partial W}{\partial x_{0}^{i}} ,
$$
in which the self-energy $W$ is given by the formula
\be \label{64}
W(x_0)=\frac{e}{2}\left( V(x)-V_{sing}(x)\right) \quad {\rm as} \quad x\rightarrow x_0 .
\ee
According to (\ref{64}), we find from formula (\ref{63}) the explicit expression
\be \label{65}
W(r_0)=-\frac{e^2wr_{0}^{2}}{2\pi (r_{0}^{2}+w^2/4)^2} .
\ee

The electrostatic self-energy (\ref{65}) can be written with the radial coordinate $l$ in the form
\be \label{66}
W(l_0)=-\frac{e^2w}{2\pi (l_{0}^{2}+w^2)} .
\ee
Result (\ref{66}) coincides with the one of Khusnutdinov and Bakhmatov \cite{khus1} found within the multipole formalism.

We now consider the general electrostatic potential (\ref{Vg0}). Since the monopole contribution is symmetric in $l$ and $l_0$,
we may use formula (\ref{64}). By adding to self-energy (\ref{66}) the monopole contribution, we find the self-energy $W_{\lambda}$
\be \label{67}
W_{\lambda}(l_0)= -\frac{e^2w}{2\pi (l_{0}^{2}+w^2)}+\frac{\lambda e^2}{2w}\left( \frac{\pi}{2}-\arctan \frac{l_0}{w}\right)^2 .
\ee

In the spacetime resulting from an identification according to the $Z_2$ symmetry, we obtain from (\ref{Z2}) the expression
of the self-energy $W_Z$ for $l_0>0$
\be \label{68}
W_Z(l_0)=-\frac{e^2w}{2\pi (l_{0}^{2}+w^2)}+\frac{e^2}{4l_0}\left[ \frac{1}{2}+\frac{1}{\pi}
\arcsin \left( \frac{-l_{0}^{2}+w^2}{l_{0}^{2}+w^2}\right) \right] .
\ee 
We point out that the self-energy $W_Z$ given by (\ref{68}) becomes large when $l_0$ is close to zero.
 
We now discuss the electrostatic self-force. It has only a radial component $f$. From self-energy (\ref{66}) corresponding to the potential $V$, we get
\be \label{70}
f(l_0)=-\frac{e^2wl_0}{\pi (l_{0}^{2}+w^2)^2} .
\ee 
We see that self-force (\ref{70}) is always attractive with respect to the throat. It vanishes at $l=0$.

For the general potential $V_{\lambda}$, we  obtain immediately from self-energy (\ref{67})
\be \label{71}
f_{\lambda}(l_0)=-\frac{e^2wl_0}{\pi (l_{0}^{2}+w^2)^2} +\frac{\lambda e^2}{l_{0}^{2}+w^2} \left( \frac{\pi}{2}-\arctan \frac{l_0}{w}\right) .
\ee
If we adopt the value $\lambda =1/\pi$ which ensures that the electric flux at infinity is $4\pi e$, we find that the 
radial component $f_{1/\pi}$ is always positive.

In the spacetime considered with $Z_2$ symmetry, we obtain from the self-energy (\ref{68}), for $l_0>0$, 
\be \label{72}
f_Z(l_0)=-\frac{e^2wl_0}{\pi (l_{0}^{2}+w^2)^2}+\frac{e^2}{8l_{0}^{2}}+\frac{e^2w}{2\pi l_0(l_{0}^{2}+w^2)}
+\frac{e^2}{4\pi l_{0}^{2}}\arcsin \left( \frac{-l_{0}^{2}+w^2}{l_{0}^{2}+w^2}\right) .
\ee
The self-force $f_Z$ is always repulsive with respect to the sphere $l=0$, edge of the spacetime. It diverges with the leading term $1/l_0$ as
$l_0\rightarrow 0$. 
 
\section{Conclusion}

We have found the physically acceptable general solution $V_{\lambda}$  to the electrostatic equation (\ref{eV}) in closed form, 
given by (\ref{Vg0}) with (\ref{Vex}).
It depends on a parameter $\lambda$ due to the fact that the electric field of the monopole solution is regular both
at the two spatial infinities. This arbitrariness is physically unsatisfactory.
We may choose $\lambda =1/\pi$
so that the electric flux $\Phi_{\lambda (\infty )}$ at the infinity $l\rightarrow \infty$ is equal to $4\pi e$. 
Infinity at $l\rightarrow \infty$ is preferred in this case and we think
that this choice is physically contrived. From this point of view, the wormhole considered with $Z_2$ symmetry
is interesting since there is unicity of the electrostatic potential $V_Z$ with obviously $\Phi_{Z (\infty)}=4\pi e$.

We have performed the calculation of the electrostatic self-force in all cases considered. For the potential $V$, the
self-force is always attractive with respect to the throat $l=0$. On the other hand for the potential $V_{\lambda}$ with $\lambda =1/\pi$,
we have found a self-force always directed toward infinity $l\rightarrow \infty$. This result is radically different from the previous one.
For the potential $V_Z$ in the spacetime considered with $Z_2$ symmetry, we have found a self-force always repulsive with respect 
to the throat, edge of the spacetime. It diverges when the point charge tends to the throat. Consequently, the charge can be held
at rest in this case only with an external force increasingly large.

\appendix

\renewcommand{\theequation}{\Alph{section}.\arabic{equation}}

\section{The Green function of the radial differential equations}

\setcounter{equation}{0}

Let the self-adjoint differential equation with a source term
\be \label{B1}
\frac{d}{dx}\left( p(x)\frac{dA}{dx}\right) +q(x)A=c\delta (x-a) ,
\ee
and the corresponding homogeneous equation
\be \label{B2}
\frac{d}{dx}\left( p(x)\frac{dA}{dx}\right) +q(x)A=0 ,
\ee
where the functions $p$, $dp/dx$ and $q$ are continuous and moreover $p$ is strictly positive on $(-\infty , \infty)$.

In the context of our present work, we assume that the homogeneous equation (\ref{B2}),
admits two regular solutions $A^{(i)}$ and $A^{(e)}$,
linearly independent, having respectively the following boundary conditions:
\be \label{B4}
\lim_{x\rightarrow -\infty} A^{(i)}(x) =0 , \quad \lim_{x\rightarrow \infty}A^{(e)}(x) =0 .
\ee
The solutions $A^{(i)}$ and $A^{(e)}$ are fixed up to a multiplicative constant.
The nonvanishing Wronskian ${\cal W}(A^{(i)},A^{(e)})$ can be easily expressed in the form
\be \label{B6}
{\cal W}( A^{(i)},A^{(e)}) (x)=\frac{k}{p(x)} ,
\ee
where $k$ is a constant.

The Green function $A(x,a)$ of the differential equation (\ref{B1}), which verifies the boundary conditions 
\be \label{B5}
\lim_{x\rightarrow -\infty} A(x,a)=0 , \quad \lim_{x\rightarrow \infty} A(x,a)=0 ,
\ee
is defined by the other following properties:
\begin{enumerate}
\renewcommand{\theenumi}{\alph{enumi}}
\item $A(x,a)$ satisfies the homogeneous equation (\ref{B2}) except at $x=a$.
\item $A(x,a)$ is continuous at $x=a$.
\item The derivative of $A(x,a)$ with respect to $x$ has a jump discontinuity of magnitude $c/p(a)$ at $x=a$.
\end{enumerate}
These conditions ensure that $A(x,a)$ satisfies the differential equation (\ref{B1}) in the sense of distributions
with the boundary conditions (\ref{B5}).

It is now possible to determine this Green function $A(x,a)$ with the help of the functions $A^{(i)}$ and $A^{(e)}$.
Consider the function $H$ defined by
\be \label{B8}
H(x,a)=\frac{c}{k}\left\{ \begin{array}{ll} A^{(i)}(x)A^{(e)}(a) \quad {\rm if} \quad x\leq a ,  \\
A^{(i)}(a)A^{(e)}(x) \quad {\rm if} \quad x\geq  a \end{array} \right. .
\ee
It is obvious that $H(x,a)$ verifies all the properties of a Green function. For simplicity, it is possible to take $k=c$ by a judicious
choice of functions $A^{(i)}$ and $A^{(e)}$ defined up to a multiplicative constant. Thus, the Green function of the differential equation (\ref{B1})
has expression (\ref{B8}) with $k=c$ that one can rewrite in the form
\be \label{B9}
A(x,a)=\theta (a-x)A^{(i)}(x)A^{(e)}(a)+\theta (x-a)A^{(i)}(a)A^{(e)}(x) ,
\ee
where $\theta$ is the Heaviside step function.

\section{Explicit electrostatic potential in higher dimensions}

\setcounter{equation}{0}

Metric (\ref{m}) can be easily generalised to describe the spacetime representing a wormhole
in $n+1$ dimensions, $n\geq 3$. We take
\be \label{Am}
ds^2=-dt^2+dl^2+( l^2+w^2) d\Omega_{n-1}^{2} ,
\ee
with $-\infty <l<\infty$, and where $d\Omega_{n-1}^{2}$ is the metric of the sphere $S^{n-1}$ described
in the coordinates $(\vartheta_1,\dots ,\vartheta_{n-2},\varphi )$.
The change of radial coordinates (\ref{ir}) expresses metric (\ref{Am}) in isotropic coordinates
\be \label{Aim}
ds^2=-dt^2+\left( 1+\frac{w^2}{4r^2}\right)^2 \left( dr^2+r^2d\Omega_{n-1}^{2}\right) 
\ee 
with $r>0$. We set the associated Cartesian coordinates $(x^i)$, $i=1,\dots ,n$.
Metric (\ref{Aim}) can be written thereby
\be \label{Acm}
ds^2=-dt^2+\left( 1+\frac{w^2}{4r^2}\right)^2\left( ( dx^1)^2+\cdots +( dx^n)^2\right) ,
\ee
where $r=\sqrt{( x^1)^2+\cdots +( x^n)^2}$.

We consider the usual Maxwell equations in a spacetime of $n+1$ dimensions. The charge density $\rho$ for a 
point charge $e$ located at $x^i=x_{0}^{i}$ has the expression
\be \label{Adc}
\rho (x^i)=\frac{e}{\sqrt{-g}}\delta^{(n)}( x^i-x_{0}^{i}) ,
\ee
where $g$ is the determinant of metric (\ref{Acm}) given by
\be \label{Ag}
\sqrt{-g}=\left(1+\frac{w^2}{4r^2}\right)^n .
\ee 
The electrostatic equation in metric (\ref{Acm}) takes the form
\be \label{AieV}
\triangle V+h(r)\frac{x^i}{r}\partial_iV=-2(n-2)\frac{\pi^{n/2}}{\Gamma (n/2)}\frac{e}{\displaystyle \left( 1+\frac{w^2}{4r^2}\right)^{n-2}} \delta^{(n)}(x^i-x_{0}^{i}) ,
\ee
where $\triangle$ is the usual Laplacian operator, and in which the function $h$ has the expression
\be \label{Ahl}
h(r)=\frac{1}{\displaystyle \left( 1+\frac{w^2}{4r^2}\right)^{n-2}}\frac{d}{dr}\left( 1+\frac{w^2}{4r^2}\right)^{n-2} =-\frac{(n-2)w^2}{2r(r^2+w^2/4)} .
\ee

In a previous work \cite{linet2} , it was shown that exact solutions to the electrostatic equation (\ref{AieV})
exist in the form
\be \label{AVex}
V(x)=g(r)g(r_0)F\left( \frac{\Gamma (x,x_0)}{k(r)k(r_0)}\right) ,
\ee
with 
\be \label{Agam}
\Gamma (x,x_0)=(x^1-x_{0}^{1})^2+\cdots +(x^n-x_{0}^{n})^2,
\ee
if the function $h$ satisfies a certain differential equation. It may be only the type II 
with $a=w/2$ in the classification of this work. Consequently, the function $h$ must satisfy 
the differential equation
\be \label{AeqII}
\frac{d}{dr}h+\frac{1}{2}h^2+\frac{n-1}{r}h-\frac{2[ -n(n-2)w^2/4-Ab^2]}{(r^2+w^2/4)^2}=0
\ee
in which the two constants $b$ et $A$ are to be fixed.

From (\ref{Ahl}), we calculate
\be \label{Aeqh}
\frac{d}{dr}h+\frac{1}{2}h^2+\frac{n-1}{r}h=-\frac{(n^2-6n+8)w ^2}{2(r^2+w^2/4)^2} .
\ee
By comparing (\ref{AeqII}) and (\ref{Aeqh}), we find the value of the constants: $b=w/2$ and $A=-4n+8$.

We are thus in the case where the electrostatic equation (\ref{AieV}) has solutions in the form (\ref{AVex}).
According to the results of the work \cite{linet2}, the functions $k$ and $g$ are expressed as
\be \label{Aekg}
k(r)=\frac{r^2+w^2/4}{w/2} , \quad g(r)=\frac{r^{n-2}}{(r^2+w^2/4)^{n-2}} ,
\ee
and the function $F$ is determined by the differential equation 
\be \label{AeqF}
4(-s^2+s)F''+(-4ns+én)F'+(-4n+8)F=0 .
\ee

As variable $s$ we have set
\be \label{Ais}
s=\frac{\Gamma (x,x_0)}{k(r)k(r_0)} .
\ee
The value $s=0$ corresponds to the position of the point charge.
We emphasize that
\be \label{A0s1}
0\leq s\leq 1 .
\ee
This comes from that fact that we can always take the particular case where the point charge is located at $\vartheta_1=0$
without lost of generality, due to the spherical symmetry.
Inequality (\ref{A0s1}) is then written
$$
w^2(r^2-2rr_0\cos \vartheta_1 +r_{0}^{2})\leq 4(r^2+w^2/4)(r_{0}^{2}+w^2/4) ,
$$
which is equivalent to the recognized inequality
$$
4r^2r_{0}^{2}+2rr_0w^2\cos \vartheta_1+w^4/4 \geq 0.
$$

We firstly find a particular solution to (\ref{AeqF})
\be \label{AFe}
F_e(s)=\frac{1}{[s(1-s)]^{n/2-1}} .
\ee
A classical method shows that the general solution to (\ref{AeqF}) is given by
\be \label{AF}
F(s)=\frac{1}{[s(1-s)]^{n/2-1}}\left\{ f_1+f_2\int_{0}^{s}[u(1-u)]^{n/2-2}du \right\}
\ee
where $f_1$ and $f_2$ are two arbitrary constants. The integral in  (\ref{AF}) can be explicitly performed
for all values of $n$.

We will choose the constants $f_1$ and $f_2$ first of all so that solution (\ref{AVex}) is regular everywhere,
except at $x=x_0$. This occurs if the function $F$, given by (\ref{AF}), has only a singularity at $s=0$. Now we have
$$
\int_{0}^{s}[u(1-u)]^{n/2-2}du -\int_{0}^{1}[u(1-u)]^{n/2-2}du \sim (1-s)^{n/2-1} \quad {\rm as}\quad s\rightarrow 1 ,
$$
therefore we take the relation
\be \label{AF2}
f_1+f_2\int_{0}^{1}[u(1-u)]^{n/2-2}du=0 .
\ee

The important point is that solution (\ref{AVex}) satisfies the electrostatic equation (\ref{AieV}) in the sense of distributions.
By noticing from (\ref{AF}) that
$$
F(s) \sim  \frac{f_1}{s^{n/2-1}} \quad {\rm as} \quad s \rightarrow 0 ,
$$
we can deduce the most divergent part of $V$ as $x\rightarrow x_0$
\be \label{AapV}
V(x) \sim  f_1\left( \frac{2}{w}\right)^{n-2}\frac{1}{(1+w^2/4r_{0}^{2})^{n/2-1}(1+w^2/4r^{2})^{n/2-1}}\frac{1}{[\Gamma (x,x_0)]^{n/2-1}}  .
\ee
By considering the source term of equation (\ref{AieV}), we take
\be \label{AF1}
f_1=e\left( \frac{w}{2}\right)^{n-2}  ,
\ee
since
$$
\triangle \frac{1}{[\Gamma (x,x_0)]^{n/2-1}} =-2(n-2)\frac{\pi^{n/2}}{\Gamma (n/2)}\delta^{(n)}(x,x_0) .
$$
So ,we obtain 
\be \label{AVf}
F(s)=e\left( \frac{w}{2}\right)^{n-2}\frac{1}{[s(1-s)]^{n/2-1}}\left\{ 1-\frac{1}{\int_{0}^{1}[u(1-u)]^{n/2-2}du}\int_{0}^{s}[u(1-u)]^{n/2-2}du \right\} .
\ee
Now, with (\ref{Aekg}), (\ref{Ais}) and (\ref{AVf}) we are in position to write from (\ref{AVex}),the potential $V(x)$ generated by a point charge at $x=x_0$.


\end{document}